\documentclass[conference]{IEEEtran} 

\usepackage{cite} % Formats citations well
\usepackage{amsmath,amssymb,amsfonts} % For math equations and symbols
\usepackage{graphicx} % For including images/diagrams
\usepackage{textcomp} % For text symbols
\usepackage{xcolor} % For color text (often used in drafts)
\usepackage{booktabs}
\usepackage{float}
\usepackage{tikz}
\usepackage{caption}
\usepackage[letterpaper, left=0.625in, right=0.625in, top=0.75in, bottom=1in]{geometry}
\renewcommand{\thetable}{\arabic{table}}
\usetikzlibrary{shapes,arrows.meta,positioning,fit,calc,backgrounds}
\begin{document}
% --- Title ---
\title{FBID: Adaptive Personalized Federated Learning for Robust Out-of-Distribution Attack Detection in IoT Networks}

% --- Author Information ---
% Note: Format changes slightly for journal vs. conference
% \author{
% \IEEEauthorblockN{An Khanh Bui\IEEEauthorrefmark{2}\IEEEauthorrefmark{3},Cong T. Nguyen\IEEEauthorrefmark{1}\IEEEauthorrefmark{2}\IEEEauthorrefmark{3}, Hoang-Anh Pham\IEEEauthorrefmark{2}\IEEEauthorrefmark{3},Dinh Thai Hoang\IEEEauthorrefmark{1}\IEEEauthorrefmark{2}\IEEEauthorrefmark{3},Diep N. Nguyen\IEEEauthorrefmark{1}\IEEEauthorrefmark{2}\IEEEauthorrefmark{3}}
% \IEEEauthorblockA{\IEEEauthorrefmark{1}UTS-HCMUT JTIRC, Ho Chi Minh City University of Technology (HCMUT), Ho Chi Minh City, 70000, Viet Nam}
% \IEEEauthorblockA{\IEEEauthorrefmark{2}Vietnam National University Ho Chi Minh City (VNU-HCM), Ho Chi Minh City, 70000, Viet Nam}
% \IEEEauthorblockA{\IEEEauthorrefmark{3}School of Electrical and Data Engineering, University of Technology Sydney, Sydney, 2007, NSW, Australia}
% }
\author{
    % --- GOM CẢ 3 TÊN VÀO 1 KHỐI N DUY NHẤT ---
    \IEEEauthorblockN{
        An Khanh Bui\textsuperscript{1,2}, Cong T. Nguyen\textsuperscript{1,2,3}, Hoang-Anh Pham\textsuperscript{1,2}, Dinh Thai Hoang\textsuperscript{3}, Diep N. Nguyen\textsuperscript{3}
    }
    
    % --- KHỐI ĐƠN VỊ CÔNG TÁC (VẪN PHẢI NẰM TRONG BLOCK A) ---
    \IEEEauthorblockA{
        \textsuperscript{1}UTS-HCMUT JTIRC, Ho Chi Minh City University of Technology (HCMUT), Ho Chi Minh City 700000, Vietnam. \\
        \textsuperscript{2}Vietnam National University Ho Chi Minh City (VNU-HCM), Ho Chi Minh City 700000, Vietnam. \\
        \textsuperscript{3}School of Electrical and Data Engineering, University of Technology Sydney, Sydney, NSW 2007, Australia.
    }
}
% \author{
%     % --- GOM CẢ 3 TÊN VÀO 1 KHỐI N DUY NHẤT ---
%     \IEEEauthorblockN{
%         Authors
%     }
    
%     % --- KHỐI ĐƠN VỊ CÔNG TÁC (VẪN PHẢI NẰM TRONG BLOCK A) ---
%     % \IEEEauthorblockA{
%     %     \textsuperscript{1}UTS-HCMUT JTIRC, Ho Chi Minh City University of Technology (HCMUT), Ho Chi Minh City 700000, Vietnam. \\
%     %     \textsuperscript{2}Vietnam National University Ho Chi Minh City (VNU-HCM), Ho Chi Minh City 700000, Vietnam. \\
%     %     \textsuperscript{3}School of Electrical and Data Engineering, University of Technology Sydney, Sydney, NSW 2007, Australia.
%     % }
% }
\maketitle

% --- Abstract ---
\begin{abstract}

Personalized Federated Learning (PFL) has emerged as a promising solution for intrusion detection in heterogeneous IoT environments, as it can improve local adaptation under highly Non-Independent and Identically Distributed (non-IID) data distributions. However, existing PFL methods often rely on client-side self-adjustment, which may lead to over-personalization and substantial degradation in out-of-distribution (OOD) attack detection. In this paper, we propose Federated Bandit Intrusion Detection (FBID), a novel adaptive PFL framework to address this limitation through server-side personalization control. In particular, FBID employs a contextual multi-armed bandit at the server to dynamically regulate each client’s local training intensity according to its observed behavior and update quality. Moreover, FBID introduces a trust-based blending mechanism to derive client-specific interpolation coefficients between the global and local models, thereby preserving global attack-detection knowledge while still allowing beneficial local specialization. Through extensive experiments on the CICIoT2023 dataset under heterogeneous client distributions and OOD stress-test settings, we show that FBID improves individual client OOD Detection Rate (DR) by up to $7.66\%$ and F1-Score (F1) by up to $5.08\%$ (relative) over the strongest stable baseline, while also improving robustness to previously unseen attack classes.
\end{abstract}

% --- Keywords ---
\begin{IEEEkeywords}
Personalized Federated Learning, Multi-Armed Bandits, Intrusion Detection, Model Robustness, IoT Security.\end{IEEEkeywords}

% --- Main Body --- 
\section{Introduction}
The rapid proliferation of Internet-of-Things (IoT) devices has fundamentally expanded the attack surface of modern networks. This expansion makes intrusion detection a critical component of IoT security systems~\cite{Moustafa_2023, Li_2022}.
 Traditional centralized intrusion detection systems (IDS) require the aggregation of large volumes of network traffic at a central server, which raises serious concerns regarding data privacy, communication overhead, and regulatory compliance. Federated Learning (FL) has recently emerged as a promising alternative by enabling distributed model training across IoT devices while keeping raw traffic data local. As a result, FL-based IDS frameworks have attracted significant attention as a privacy-preserving and scalable solution for collaborative threat detection~\cite{Arisdakessian_2023}.

Despite these advantages, the effectiveness of conventional FL is severely limited in realistic IoT deployments due to extreme statistical heterogeneity~\cite{Nguyen_2021}. In particular, IoT devices differ widely in functionality, traffic volume, protocol usage, and exposure to attack types, which leads to highly Non-Independent and Identically Distributed (non-IID) data distributions across clients.
 Under such conditions, standard FL aggregation methods, such as FedAvg~\cite{mcmahan2017communication}, often converge to suboptimal global models that fail to capture client-specific characteristics and exhibit poor generalization under distribution shift, as recently highlighted in PFL-based IoT intrusion detection.

To address data heterogeneity, Personalized Federated Learning (PFL) has been proposed across diverse IoT domains~\cite{wu2020personalized}, allowing each client to maintain a model that balances global knowledge with local specialization. PFL approaches, such as Ditto~\cite{li2021ditto}, APFL~\cite{deng2020adaptive}, and FedALA~\cite{fedala_ref}, introduce personalization through regularization or adaptive blending between global and local models. While these methods improve local performance, they rely exclusively on client-side self-adjustment. When personalization is driven solely by local optimization objectives, clients with skewed or low-diversity data may progressively bias their models toward local optima. The resulting loss of global knowledge reduces robustness, especially against out-of-distribution (OOD) threats. Such threats are critical in IoT contexts as they often represent novel zero-day attacks or evolving malicious signatures that bypass localized detection patterns, potentially compromising the entire network. The objective of this work is to mitigate this effect while retaining effective local adaptation.

In this paper, we develop Federated Bandit Intrusion Detection (FBID), a PFL framework for heterogeneous IoT intrusion detection that introduces server-side supervision of personalization to specifically overcome the OOD generalization problem. Instead of leaving adaptation entirely to client-side local objectives, FBID uses global feedback at the server to regulate both how deeply each client trains locally and how strongly its personalized model is allowed to deviate from the global model. Specifically, FBID employs a server-side contextual multi-armed bandit to assign client-specific local training budgets. Moreover, a trust-based blending mechanism is developed to derive client-specific interpolation parameters between the global and local models, thereby increasing robustness against OOD attacks while still allowing beneficial local specialization. In addition, FBID evaluates client updates against a compact server-side validation set to support server-side estimation of update quality. Extensive experiments are conducted on the CICIoT2023 dataset under heterogeneous client distributions and explicit OOD attack settings. The results show that FBID improves OOD intrusion detection performance across diverse clients, achieving relative gains in individual client performance of up to $7.66\%$ in Detection Rate (DR) and $5.08\%$ in F1-Score (F1) over state-of-the-art baselines, while mitigating the OOD vulnerabilities observed in client-driven adaptive methods.
\section{System Model and Problem Formulation}
\subsection{System Overview}
\label{subsec:system_overview}

We consider an FL-based intrusion detection system operating in a heterogeneous IoT environment. The system comprises a single central coordination server and a set $\mathcal{K} = \{1, \dots, K\}$ of IoT client nodes. Each client $k \in \mathcal{K}$ represents a network edge device, such as an IoT gateway, a smart hub, or a compute-capable edge node, tasked with monitoring local network traffic for malicious activities. In this distributed architecture, each client $k$ continuously collects a local dataset $\mathcal{D}_k = \{(\mathbf{f}_{k,i}, y_{k,i})\}_{i=1}^{n_k}$ of network flows, where $\mathbf{f}_{k,i} \in \mathbb{R}^d$ represents the network flow feature vector and $y_{k,i}$ denotes the corresponding intrusion label.

To collaboratively build an intrusion detection model without compromising raw data privacy, the clients participate in a standard FL protocol coordinated by the central server. The server maintains a global model $w^G$ and facilitates periodic communication rounds. In a typical round, the server broadcasts the global model to a subset of available clients. Each selected client performs local training on its private dataset $\mathcal{D}_k$ to update the model parameters, and subsequently transmits only the model updates back to the server. Furthermore, the central server maintains a compact validation dataset $\mathcal{D}_{val}$ curated from public threat intelligence corpora. This dataset is strictly isolated from the clients' private local data and serves as a neutral baseline to evaluate the generalization capacity of incoming client updates.

\begin{figure}[!]
\centering
\includegraphics[width=0.85\columnwidth]{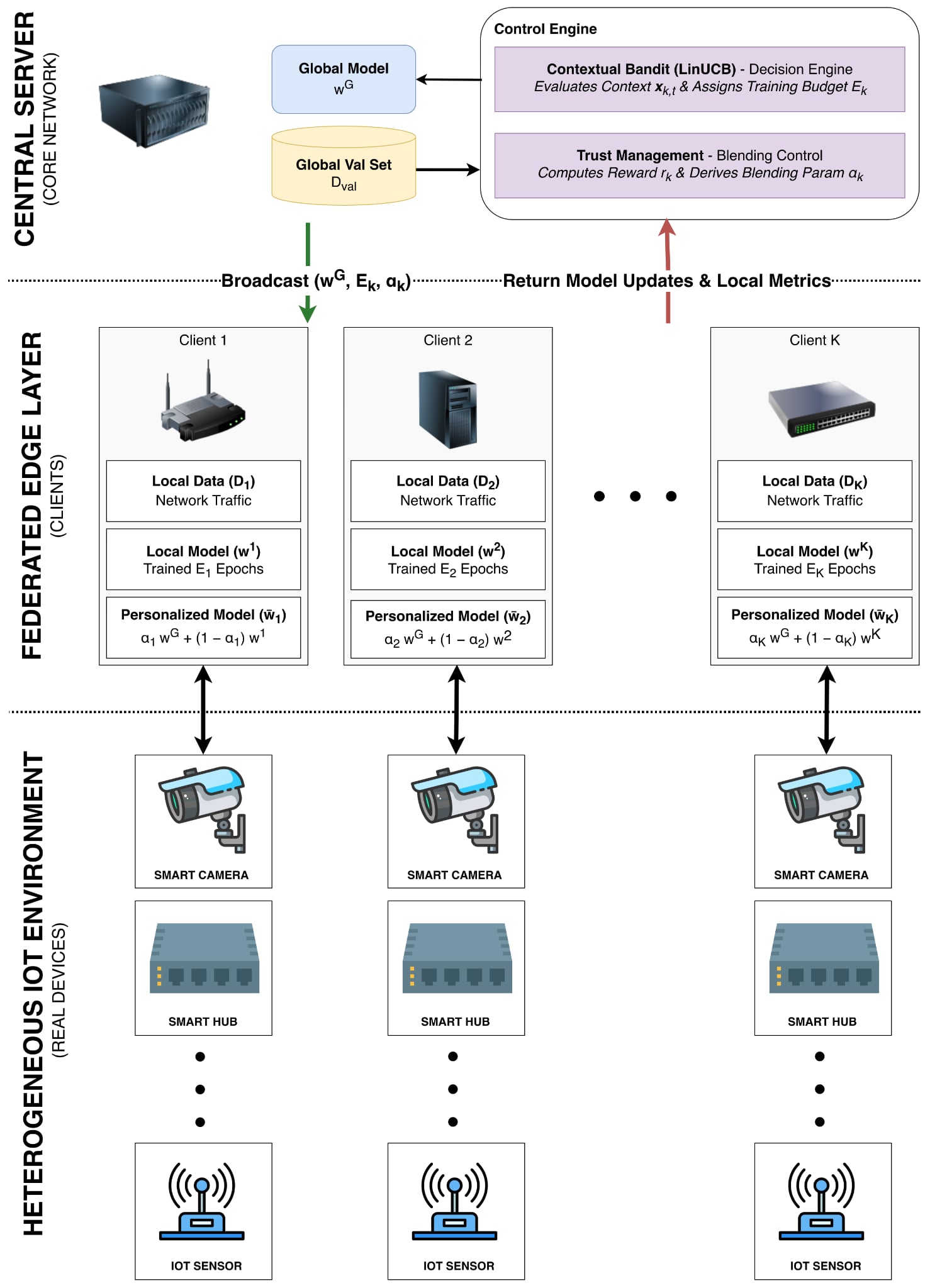}
\caption{The FBID framework overview.}
\label{fig:system_overview}
\end{figure}

\subsection{Problem Formulation}

Due to the diversity of IoT deployments and threat exposures, the local data distributions are inherently non-IID, i.e., $\mathcal{P}_k \neq \mathcal{P}_{k'}$ for $k \neq k'$. The system follows the protocol described in Section II-A, where the server coordinates iterative communication rounds to refine the global model $w^G$ and utilizes the validation dataset $\mathcal{D}_{\mathrm{val}}$ to assess update quality.

The objective of FBID is to learn a set of personalized models $\{\bar{w}_k\}_{k=1}^K$ that balance local specialization and global generalization. For each client $k$, the personalized model is defined as a convex combination of the global model $w^G$ and a local model $w^k$:
\begin{equation}
\bar{w}_k = \alpha_k w^G + (1 - \alpha_k) w^k,
\label{eq:blending}
\end{equation}
where $\alpha_k \in [0, 1]$ is the personalization coefficient, which is enforced to remain in $[0,1]$ by the server-side trust mechanism described in Section III-B. The global training objective is formulated as:
\begin{equation}
\min_{\{w^k\}_{k=1}^K,\, w^G} \sum_{k=1}^K \frac{n_k}{N} \mathcal{L}_k(\bar{w}_k),
\label{eq:objective}
\end{equation}
where $N = \sum_{k=1}^K n_k$ is the total number of training samples, $\mathcal{L}_k(\cdot)$ denotes the local empirical loss on $\mathcal{D}_k$, and $\alpha_k$ are fixed server-assigned parameters rather than optimization variables. Here, $w^k$ implicitly depends on the server-assigned budget $E_k$, as the local model is obtained by running $E_k$ epochs of SGD from $w^G$. In conventional PFL approaches, the coefficients $\alpha_k$ are optimized locally at each client~\cite{deng2020adaptive, hanzely2020federated}. However, under highly non-IID conditions, treating $\alpha_k$ as a purely local optimization variable can lead to degenerate solutions~\cite{deng2020adaptive, luo2022apple}, where clients favor $\alpha_k \rightarrow 0$ to overfit their local data~\cite{hanzely2020federated, deng2020adaptive}. This behavior reduces reliance on the global model and degrades the ability to detect previously unseen attacks. To address this limitation, FBID introduces a server-side control mechanism that adaptively adjusts $\alpha_k$, ensuring an appropriate balance between personalization and global knowledge, thereby preserving robustness against OOD threats.

% =============================================================
\section{The Proposed FBID Framework}
% =============================================================

The FBID framework comprises three coordinated components: (i) a \textit{Server-Side Contextual Bandit} that adaptively assigns training budgets, (ii) an \textit{Adaptive Trust-Based Blending} 
mechanism that derives client-specific interpolation parameters, and (iii) a \textit{Global Aggregation} step. Although LinUCB, exponential moving averages, and convex interpolation are individually established techniques, their naive composition introduces a fundamental credit assignment problem: a single validation reward cannot disambiguate whether poor performance stems from insufficient local training or excessive deviation from the global model. FBID resolves this by \textit{functionally} decoupling optimization from interpolation: the bandit explicitly controls the optimization trajectory (epoch budget $E_k$) under a drift-penalized reward, while the population-calibrated trust mechanism independently governs the degree of model deviation ($\alpha_k$). This asymmetric design prevents the runaway over-personalization characteristic of naive compositions under highly non-IID conditions.

\subsection{Server-Side Contextual Bandit Control}
The innovation of \textit{FBID} is moving the decision-making intelligence to the server-side. We develop a mechanism based on the Contextual Multi-Armed Bandit, specifically utilizing the LinUCB algorithm~\cite{linucb_ref}, to manage the exploration-exploitation trade-off involved in local optimization.

\subsubsection{Context Space $\mathbf{x}_k$}
At each round $t$, the server constructs a 6-dimensional context vector for client $k$:
$\mathbf{x}_{k,t} = [n_k/N,\; \text{DR}_k,\; \text{FPR}_k,\; \ell_{val,k},\; \tau_k,\; t/T] \in \mathbb{R}^6$.
The features are: relative sample size $n_k/N \in [0,1]$, local Detection Rate and FPR both in $[0,1]$, the local validation loss $\ell_{val,k}$, measured as raw cross-entropy and left unstandardized, current trust score $\tau_k \in [0,1]$, and normalized training progress $t/T \in [0,1]$. Most context features are naturally bounded in $[0,1]$. While the validation loss is not standardized, the LinUCB regularization $A_k^{(0)} = I$ provides implicit scale robustness, and empirically this mild non-uniformity did not degrade bandit convergence.

\subsubsection{Action and Reward}
The action space is defined as a discrete set $\mathcal{A} = \{1, 2, 3, 5\}$ of local epochs, representing the training budget allocated to each client per round. The chosen values form distinct computational tiers, namely minimal (1 epoch), moderate (2 to 3 epochs), and extended (5 epochs) of local training. The upper bound of 5 is consistent with the FL literature, where excessively deep local training has been shown to exacerbate client drift under non-IID conditions~\cite{fedeff_ref}, while adaptive epoch adjustment via a discrete action space has been validated in heterogeneous IoT FL settings~\cite{fedddrl_ref}. For nodes far from convergence, the AUC improvement on $\mathcal{D}_{\mathrm{val}}$ dominates the drift penalty in $r_k$, while for nodes near convergence, diminishing AUC returns allow the drift penalty to dominate. After client execution, the server computes a reward $r_k$:
\begin{equation}
r_k = \lambda_1 \delta_{\text{AUC}}^{(k)} + \lambda_2 F1_k - \lambda_3 \|\Delta_k / E_k\|_2,
\label{eq:reward}
\end{equation}
where $F1_k$ is the local F1-Score on local data, $\|\Delta_k / E_k\|_2$ is the L2 norm of the per-epoch weight update, normalized to make drift comparable across epoch assignments, and $\delta_{\text{AUC}}^{(k)}$ represents a per-client AUC contribution signal:
\begin{equation}
\delta_{\text{AUC}}^{(k)} = \text{AUC}(w^G + \Delta_k/E_k) - \text{AUC}(w^G).
\label{eq:delta_auc}
\end{equation}
This is computed by temporarily applying client $k$'s scaled update to the pre-aggregation global model $w^G$, evaluating AUC on $\mathcal{D}_{val}$, and then restoring $w^G$ to its original state (e.g., via temporary model shadowing) to avoid accumulation of floating-point drift. While this requires $K$ forward passes over $\mathcal{D}_{val}$ per round, it provides each client a distinct, individually attributable reward signal.

\subsection{Adaptive Trust-Based Blending}
The final personalized model $\bar{w}_k$ is computed through a dynamic interpolation of the global model $w^G$ and the local model $w^k$:
\begin{equation}
\bar{w}_k = \alpha_k w^G + (1 - \alpha_k) w^k.
\end{equation}
The blending parameter $\alpha_k$ is derived from a trust score $\tau_k \in [0,1]$ representing trust in the client's local model, where high trust (in the local model) indicates greater local autonomy (lower $\alpha_k$), and low trust pulls the client toward $w^G$. The trust score is updated each round via an exponential moving average of the sigmoid-transformed, z-score-normalized reward $\hat{r}_k$:
\begin{equation}
\tau_k \leftarrow \beta \tau_k + (1-\beta) \sigma(\hat{r}_k),
\label{eq:trust}
\end{equation}
\begin{equation}
\alpha_k = 1 - \tau_k,
\label{eq:blending_trust}
\end{equation}
where $\hat{r}_k$ is $r_k$ z-score-normalized across all selected clients in each communication round, and $\beta \in (0, 1)$ is a trust decay factor controlling the rate at which past rewards are discounted. Normalizing the reward prevents long-term shifts in the absolute reward scale from inflating or deflating trust scores over training. The sigmoid transformation constrains the trust score strictly within $(0, 1)$, preventing complete local-model collapse. High-trust clients ($\tau_k \rightarrow 1$) receive small $\alpha_k$, granting local autonomy, while low-trust clients are pulled toward $w^G$ to utilize more stable global knowledge.

\begin{figure}[!]
\centering
\begin{minipage}{\columnwidth}
\hrule height 1pt
\vspace{2pt}
\textbf{Algorithm 1:} FBID Controller
\vspace{2pt}
\hrule
\vspace{2pt}
{\footnotesize
\textbf{Require:} Rounds $T$, Clients $K$, Global validation set $\mathcal{D}_{val}$ \\
\textbf{Ensure:} Personalized models $\{\bar{w}_k\}_{k=1}^K$ \\[2pt]
\textbf{1:} Initialize $w^G$, set $\tau_k \leftarrow 0.5$ for all $k$ \\[1pt]
\textbf{2:} \textbf{for} each round $t = 1, \dots, T$ \textbf{do} \\
\textbf{3:} \hspace{1em}\textbf{for} each selected client $k$ \textbf{do} \\
\textbf{4:} \hspace{2em}Compute context vector $\mathbf{x}_{k,t}$\\
\textbf{5:} \hspace{2em}Select $E_k \leftarrow \text{LinUCB}(\mathbf{x}_{k,t})$, set $\alpha_k \leftarrow 1 - \tau_k$ \\
\textbf{6:} \hspace{2em}Send $(w^G,\, E_k,\, \alpha_k)$ to client $k$ \\
\textbf{7:} \hspace{2em}Client trains $E_k$ epochs, returns $w^k$ \\
\textbf{8:} \hspace{2em}$r_k \leftarrow \lambda_1 \delta_{\text{AUC}}^{(k)} + \lambda_2\text{F1}_k - \lambda_3\|\Delta_k/E_k\|_2$ \\
\textbf{9:} \hspace{1em}\textbf{end for} \\
\textbf{10:} \hspace{1em}$\hat{r}_k \leftarrow$ z-normalize $\{r_k\}$ across selected clients \\
\textbf{11:} \hspace{1em}\textbf{for} each selected client $k$ \textbf{do} \\
\textbf{12:} \hspace{2em}Update LinUCB with $(\mathbf{x}_{k,t}, E_k, \hat{r}_k)$ \\
\textbf{13:} \hspace{2em}$\tau_k \leftarrow \beta\tau_k + (1-\beta)\sigma(\hat{r}_k)$ \\
\textbf{14:} \hspace{1em}\textbf{end for} \\
\textbf{15:} \hspace{1em}$w^G \leftarrow w^G + \displaystyle\sum_{k} \frac{\tau_k n_k}{\sum_j \tau_j n_j} \cdot \frac{\Delta_k}{E_k}$ \\
\textbf{16:} \hspace{1em}\textbf{for} each selected client $k$ \textbf{do} \\
\textbf{17:} \hspace{2em}$\bar{w}_k \leftarrow \alpha_k w^G + (1-\alpha_k)w^k$ \\
\textbf{18:} \hspace{1em}\textbf{end for} \\
\textbf{19:} \textbf{end for}
}
\vspace{4pt}
\hrule height 1pt
\end{minipage}
\vspace{10pt}
% Uncomment the next line if you want \ref{alg:fbid} to work correctly in text
% \caption{FBID Controller Algorithm}
\label{alg:fbid}
\end{figure}

Algorithm~1 outlines the operational workflow of the FBID controller through a three-phase structure within each communication round. In the first phase, the server computes each client's context vector $\mathbf{x}_{k,t}$, selects a training budget $E_k$ via LinUCB, derives the current blending coefficient $\alpha_k = 1 - \tau_k$ from the trust score accumulated in the previous round, and dispatches $(w^G, E_k, \alpha_k)$ to the client. Each client trains for $E_k$ epochs and the server computes a per-client reward $r_k$. In the second phase, z-score normalization is applied to $r_k$ to produce a population-calibrated score $\hat{r}_k$, the normalized reward is then used both to update the LinUCB policy and to update the trust score $\tau_k$ via an exponential moving average. This phase is necessarily deferred until all rewards are observed, as z-normalization requires the full cohort's performance to establish a meaningful relative baseline. In the third phase, the global model is updated via trust-weighted aggregation, where each client's contribution is weighted by its trust score $\tau_k$ and dataset size $n_k$. The personalized model $\bar{w}_k$ is then formed using the $\alpha_k$ dispatched at the start of the round, the updated trust score $\tau_k$ will govern $\alpha_k$ in the subsequent round. Regarding initialization, setting $\tau_k = 0.5$ for all clients establishes a neutral prior: in round $t = 1$, this yields uniform $\alpha_k = 0.5$ and uniform exploration budgets across all clients, with heterogeneous allocation emerging only after the LinUCB controller accumulates sufficient context statistics to favor informed exploitation.
% =============================================================
\section{Performance Evaluation}
% =============================================================

\subsection{Experiment Setup}

\subsubsection{Dataset Description}
We evaluate the FBID framework using the CICIoT2023 dataset~\cite{ciciot2023}, a large-scale, high-fidelity network security corpus designed for benchmarking machine learning models in IoT environments. The dataset contains traffic profiles for 33 distinct attack types and benign traffic, captured across a diverse range of IoT protocols. For our experiments, we utilize the 46-feature tabular representation of the network flows, which includes flow durations, packet counts, and statistical features of traffic bursts.

In a practical IoT or edge-computing environment, data collected by participating devices is rarely IID. Devices possess varying storage capacities and face entirely different threat landscapes depending on their network location, purpose, and exposure. To simulate this heterogeneity, we partition the training data across $K=10$ distinct nodes using a severely non-IID strategy based on the CICIoT2023 dataset. The data is divided into four distinct profile categories.

\begin{itemize}
    \setlength{\itemsep}{0pt}
    \item \textbf{Global Gateways (Nodes 1--3):} 120,000 samples ($85\%$ benign) focused on volumetric DoS/DDoS (TCP, SYN, UDP, ICMP).
    \item \textbf{Internal IoT Hubs (Nodes 4--6):} 64,000 samples (95\% benign) featuring lateral threats (DNS/ARP Spoofing, Brute Force).
    \item \textbf{Internal Subnet Controllers (Nodes 7--8):} 16,000 purely benign samples for false positive verification.
    \item \textbf{Edge Gateways (Nodes 9--10):} 16,000 samples (90\% benign) with a mix of DoS/DDoS and slow-rate attacks.
\end{itemize}

To evaluate FBID's robustness under distributional shifts, we conduct an Out-of-Distribution (OOD) stress test using 20,263 unseen samples across 21 traffic classes (detailed in Table~\ref{tab:ood_dataset}). To rigorously test sensitivity to rare cyberattacks and prevent artificial performance inflation, this OOD set introduces novel attack types that are entirely absent from the clients' training data. Evaluation primarily relies on the OOD Detection Rate, supported by the F1-Score and False Positive Rate (FPR).

\begin{table}[!]
\caption{Detailed distribution of global stress test dataset.}
\label{tab:ood_dataset}
\begin{center}
\renewcommand{\arraystretch}{0.9}
\begin{tabular}{@{} l r | l r @{}}
\toprule
\textbf{Attack Class} & \textbf{Count} & \textbf{Attack Class} & \textbf{Count} \\
\midrule
BenignTraffic       & 8,433 & Recon-OSScan         & 422 \\
DoS-TCP\_Flood      & 1,687 & Recon-PortScan       & 422 \\
DoS-SYN\_Flood      & 1,687 & Mirai-greeth\_flood  & 422 \\
DDoS-UDP\_Flood     & 1,265 & VulnerabilityScan    & 422 \\
DDoS-SlowLoris      & 1,265 & SqlInjection         & 173 \\
DDoS-ICMP\_Flood    &   844 & CommandInjection     & 180 \\
Mirai-udpplain      &   844 & BrowserHijacking     & 169 \\
Recon-HostDiscovery &   496 & XSS                  & 124 \\
MITM-ArpSpoofing    &   422 & Backdoor\_Malware    & 104 \\
DictionaryBruteForce&   422 & Uploading\_Attack    &  38 \\
DNS\_Spoofing       &   422 & \textbf{Total Samples}& \textbf{20,263} \\
\bottomrule
\end{tabular}
\end{center}
\end{table}

\subsubsection{Implementation Details and Hyperparameters}
Experiments are implemented in PyTorch using PFLlib~\cite{pfllib_ref} on CPUs. Our intrusion detection model, \textit{IDSNet}, is a 3-layer MLP (64, 32, 16 units) with ReLU activations. Federated training runs for 200 communication rounds across 10 clients. Local optimization utilizes SGD (batch size 512, learning rate 0.005, decay $\gamma = 0.99$). All metrics are averaged across three independent seeds. For FBID, reward weights are $\lambda_1 = 1.0$, $\lambda_2 = 0.5$, and $\lambda_3 = 0.01$, with a trust decay $\beta = 0.85$. Because $\delta_{\text{AUC}}^{(k)}$ and local F1 are bounded [0, 1] probabilities, whereas the unconstrained weight update norm $\|\Delta_k/E_k\|_2$ is magnitudes larger, $\lambda_3$ functions as a normalizing constant to equalize the loss geometry and prevent the drift penalty from suppressing exploration. We evaluate FBID against four baselines: FedALA~\cite{fedala_ref} (global-optimization PFL), APFL~\cite{deng2020adaptive} (local-mixing PFL), Ditto~\cite{li2021ditto} (regularization-based PFL), and CBC (Continuous Bandit Controller), an ablation baseline. CBC is a client-driven adaptive PFL method that locally optimizes the blending parameter $\alpha_k$ using SPSA~\cite{spall1992multivariate}. Clients maintain a persistent local model $w^k$ alongside the global model $w^G$, updating $\alpha_k$ via $\alpha_{k}^{(t+1)} = \alpha_k^{(t)} + \eta [R(\alpha + \delta) - R(\alpha - \delta)]/2\delta$. Here, $R$ is a validation reward (F1-Score), while step size $\eta$ and perturbation $\delta$ anneal over rounds. The final personalized model is $\bar{w}_k = \alpha_k w^G + (1 - \alpha_k) w^k$.

% \subsubsection{Baseline Algorithms}
% \label{sec:baselines}

\begin{table}[!]
\caption{Aggregated detection performance.}
\label{tab:id_metrics}
\resizebox{\columnwidth}{!}{

\begin{tabular}{|l|cc|cc|}
\toprule
& \multicolumn{2}{c|}{\textbf{In-Distribution (ID)}} 
& \multicolumn{2}{c|}{\textbf{Out-of-Distribution (OOD)}} \\
\midrule
\textbf{Algorithm} & \textbf{F1} & \textbf{DR} & \textbf{F1} & \textbf{DR} \\

\midrule

\textbf{FBID} & 0.899 $\pm$ 0.018 & 0.821 $\pm$ 0.030 & \textbf{0.680} $\pm$ 0.066 & \textbf{0.518} $\pm$ 0.074 \\

CBC & 0.896 $\pm$ 0.019 & 0.816 $\pm$ 0.031 & 0.655 $\pm$ 0.070 & 0.490 $\pm$ 0.076 \\

FedALA & 0.896 $\pm$ 0.017 & 0.816 $\pm$ 0.029 & 0.660 $\pm$ 0.056 & 0.494 $\pm$ 0.061 \\

Ditto & 0.899 $\pm$ 0.017 & 0.819 $\pm$ 0.028 & 0.548 $\pm$ 0.041 & 0.379 $\pm$ 0.039 \\

APFL & \textbf{0.901} $\pm$ 0.019 & \textbf{0.823} $\pm$ 0.032 & 0.524 $\pm$ 0.061 & 0.356 $\pm$ 0.054 \\

\bottomrule
\end{tabular}
}
\end{table}
\subsection{Results and Discussion }

\subsubsection{Aggregated Performance Comparison}
\label{sec:average}
As shown in Table~\ref{tab:id_metrics}, all methods exhibit comparable In-Distribution (ID) performance, with pooled F1 scores spanning a narrow $0.5\%$ range. APFL attains the highest F1 ($0.901$) and detection rate (DR) ($0.823$). Configured with a fixed blending coefficient $\alpha = 0.5$ in our implementation, APFL provides an equal weighting of global and local models that proves optimal under balanced data. FBID and Ditto perform comparably (F1 = $0.899$). Furthermore, all frameworks maintain near-zero false positive rates (FPR), satisfying a critical prerequisite for minimizing false alarms in operational deployments.

The OOD evaluation, however, reveals a substantially different picture (Table~\ref{tab:id_metrics}). FBID achieves the highest F1 ($0.680$) and DR ($0.518$) among methods that maintain stability across all client clusters, representing an absolute improvement of $2.0\%$ in F1 and $2.4\%$ in DR over the next-best stable baseline, FedALA (F1 $= 0.660$, DR $= 0.494$). Notably, APFL's fixed blending coefficient $\alpha = 0.5$, despite its in-distribution advantage, degrades significantly on OOD data (F1 $= 0.524$, DR $= 0.356$), demonstrating that a static mixing coefficient cannot adequately preserve global attack knowledge under distributional shift. FBID also outperforms the CBC ablation by $2.5\%$ in F1 and $2.8\%$ in DR, confirming the benefit of server-side global oversight over unconstrained client-driven adaptation. Ditto similarly fails on benign-heavy clients. These results show in-distribution performance fails to guarantee deployment robustness, making server-side adaptive personalization essential for reliable OOD detection.

\subsubsection{Per-client Analysis}
\label{sec:failure}

As shown in the per-client evaluations in Table \ref{tab:client_stability_combined}, the proposed FBID framework generally outperforms all stable baseline methods across clients. The only exception is observed with adaptive APFL at Clients 1--3, where APFL achieves slightly higher OOD F1 and DR than FBID. However, this marginal gain comes at a severe cost: APFL collapses completely at Clients 4--6 (Internal IoT Hubs with 95\% benign traffic), missing almost 100\% of attacks, a critical vulnerability for any IDS. 

FBID consistently outperforms all other baseline approaches. Compared to FedALA, FBID improves F1 and DR by up to $3.20\%$ and $3.70\%$, respectively. Compared to CBC, FBID achieves up to $2.70\%$ higher F1 and $3.00\%$ higher DR. The regularization-based Ditto framework fails critically at Clients 7 and 8 (benign-only nodes), reducing their DR virtually to zero, while underperforming FBID at all other clients. These results confirm that server-side control of personalization helps FBID maintain more stable OOD performance across diverse client profiles than client-driven adaptive methods.

\begin{table*}[!t]
\caption{Per-client OOD performance.}
\label{tab:client_stability_combined}
\centering
\footnotesize
\setlength{\tabcolsep}{4pt}
\resizebox{2\columnwidth}{!}{
\renewcommand{\arraystretch}{1.3}
\begin{tabular}{@{} l | c c | c c | c c | c c | c c @{}}
\hline
 & \multicolumn{2}{c|}{\textbf{FBID}} & \multicolumn{2}{c|}{\textbf{FedALA}} & \multicolumn{2}{c|}{\textbf{APFL}} & \multicolumn{2}{c|}{\textbf{CBC}} & \multicolumn{2}{c}{\textbf{Ditto}} \\
\textbf{Client} & \textbf{F1} & \textbf{DR} & \textbf{F1} & \textbf{DR} & \textbf{F1} & \textbf{DR} & \textbf{F1} & \textbf{DR} & \textbf{F1} & \textbf{DR} \\
\hline
    C1 & 0.679$\pm$0.054 & 0.517$\pm$0.060 & 0.670$\pm$0.052 & 0.507$\pm$0.057 & $\mathbf{0.695}$$\pm$0.059 & $\mathbf{0.536}$$\pm$0.067 & 0.662$\pm$0.060 & 0.498$\pm$0.066 & 0.679$\pm$0.053 & 0.516$\pm$0.060 \\
    C2 & 0.679$\pm$0.054 & 0.517$\pm$0.060 & 0.670$\pm$0.052 & 0.507$\pm$0.057 & $\mathbf{0.698}$$\pm$0.059 & $\mathbf{0.539}$$\pm$0.068 & 0.662$\pm$0.060 & 0.498$\pm$0.065 & 0.679$\pm$0.054 & 0.517$\pm$0.061 \\
    C3 & 0.679$\pm$0.053 & 0.516$\pm$0.060 & 0.670$\pm$0.051 & 0.506$\pm$0.057 & $\mathbf{0.694}$$\pm$0.058 & $\mathbf{0.535}$$\pm$0.066 & 0.657$\pm$0.059 & 0.493$\pm$0.064 & 0.677$\pm$0.053 & 0.515$\pm$0.059 \\
    \hline
    C4 & $\mathbf{0.682}$$\pm$0.054 & $\mathbf{0.520}$$\pm$0.061 & 0.650$\pm$0.041 & 0.483$\pm$0.045 & 0.001$\pm$0.001 & 0.000$\pm$0.001 & 0.655$\pm$0.058 & 0.490$\pm$0.062 & 0.547$\pm$0.073 & 0.380$\pm$0.067 \\
    C5 & $\mathbf{0.682}$$\pm$0.054 & $\mathbf{0.520}$$\pm$0.061 & 0.650$\pm$0.041 & 0.483$\pm$0.044 & 0.001$\pm$0.001 & 0.000$\pm$0.001 & 0.654$\pm$0.057 & 0.489$\pm$0.062 & 0.548$\pm$0.073 & 0.381$\pm$0.068 \\
    C6 & $\mathbf{0.682}$$\pm$0.054 & $\mathbf{0.520}$$\pm$0.061 & 0.649$\pm$0.041 & 0.482$\pm$0.044 & 0.001$\pm$0.001 & 0.000$\pm$0.000 & 0.654$\pm$0.057 & 0.488$\pm$0.061 & 0.548$\pm$0.073 & 0.381$\pm$0.068 \\
    \hline
    C7 & $\mathbf{0.678}$$\pm$0.053 & $\mathbf{0.516}$$\pm$0.059 & 0.657$\pm$0.044 & 0.491$\pm$0.048 & 0.653$\pm$0.056 & 0.487$\pm$0.061 & 0.650$\pm$0.055 & 0.484$\pm$0.059 & 0.108$\pm$0.089 & 0.059$\pm$0.050 \\
    C8 & $\mathbf{0.678}$$\pm$0.054 & $\mathbf{0.516}$$\pm$0.060 & 0.657$\pm$0.044 & 0.491$\pm$0.048 & 0.653$\pm$0.056 & 0.487$\pm$0.061 & 0.650$\pm$0.055 & 0.484$\pm$0.059 & 0.115$\pm$0.095 & 0.064$\pm$0.054 \\
    \hline
    C9 & $\mathbf{0.680}$$\pm$0.054 & $\mathbf{0.517}$$\pm$0.060 & 0.660$\pm$0.046 & 0.494$\pm$0.051 & 0.654$\pm$0.057 & 0.489$\pm$0.062 & 0.653$\pm$0.057 & 0.488$\pm$0.061 & 0.654$\pm$0.003 & 0.487$\pm$0.003 \\
    C10 & $\mathbf{0.680}$$\pm$0.053 & $\mathbf{0.517}$$\pm$0.060 & 0.661$\pm$0.046 & 0.496$\pm$0.051 & 0.654$\pm$0.058 & 0.489$\pm$0.062 & 0.653$\pm$0.057 & 0.487$\pm$0.061 & 0.652$\pm$0.004 & 0.484$\pm$0.004 \\
\hline
\end{tabular}
}
\end{table*}

\subsubsection{Analysis of Adaptive Blending Dynamics}
\label{sec:blending}
\begin{figure}[!]
\centerline{\includegraphics[width=0.9\columnwidth]{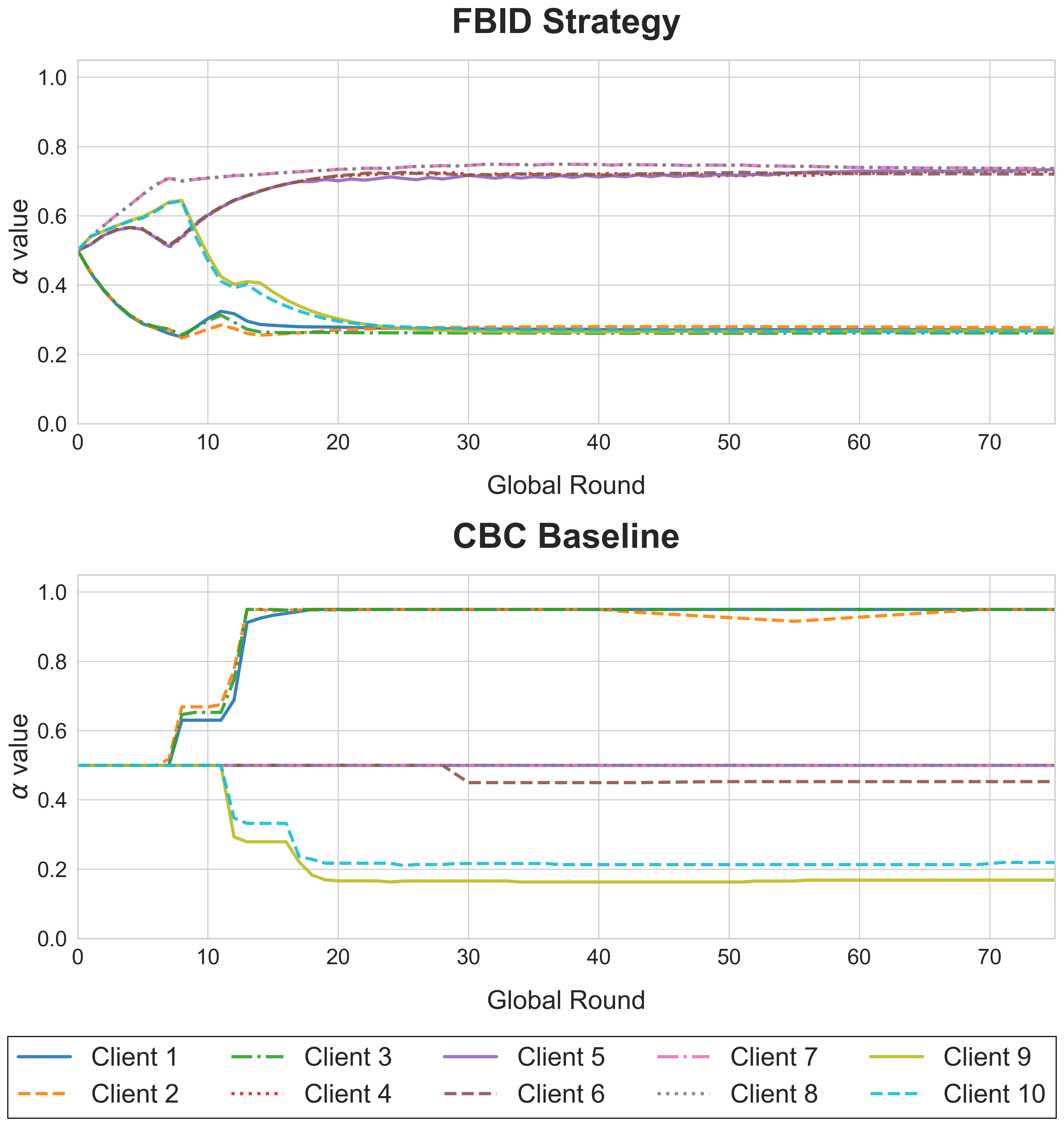}}
\caption{Trajectory of $\alpha$ over the first 75 communication rounds.}
\label{fig:fbid_cbc_dynamics}
\end{figure}

The effectiveness of FBID depends on whether the server-side controller can meaningfully differentiate between reliable and unreliable client updates over time. In FBID, this differentiation is reflected in the evolution of the blending parameter $\alpha_k$, which is derived from the reward-driven trust score $\tau_k$ and determines the relative influence of the global and local models in each client’s personalized model. As shown in Fig.~\ref{fig:fbid_cbc_dynamics}, the $\alpha_k$ trajectories for the FBID Strategy do not converge to a common intermediate value. Instead, they gradually separate into two distinct regimes. Reliable clients (Clients 1--3 and 9--10), whose local updates are consistently beneficial, tend to converge toward lower $\alpha_k$ values (approximately 0.27), indicating that the server permits stronger local specialization for these nodes. In contrast, clients with more skewed or benign-heavy local datasets (Clients 4--8) converge toward higher $\alpha_k$ values (approximately 0.73), which keeps their personalized models naturally closer to the global model to preserve robustness. This behavior suggests that FBID adaptively adjusts personalization strength according to observed client update quality, rather than applying a uniform personalization policy across the federation.

Unlike FBID, CBC converges to suboptimal $\alpha_k$ values that ignore underlying data distributions, leading to two distinct failure modes. First, in nodes with highly skewed data distributions (Clients 4--8, predominantly benign traffic), CBC drives $\alpha_k$ too low, causing severe over-personalization. Due to the overfitting to local benign data and the discarding of global attack knowledge, OOD detection drops significantly (e.g., Client 4 under CBC yields F1 = 0.655, DR = 0.490, compared with FBID F1 = 0.682, DR = 0.520). Conversely, for information-rich gateway nodes (Clients 1--3), CBC pushes $\alpha_k$ near 0.95. This is counterintuitive because nodes with high-quality, diverse local data are typically the best candidates for local specialization. However, CBC's unconstrained local optimization leads to a 'lazy' over-reliance on the global model, suppressing the beneficial specialization that FBID correctly identifies with $\alpha_k \approx 0.27$. Unlike CBC's unconstrained client optimization, FBID's server control aligns $\alpha_k$ with data quality, ensuring superior OOD robustness.

\subsubsection{Further Discussion and System Overhead}
FBID requires a compact server-side validation set $\mathcal{D}_{\mathrm{val}}$ to assess the usefulness of client updates. This does not alter the standard FL privacy model, since raw client traffic remains local and $\mathcal{D}_{\mathrm{val}}$ is used only for server-side evaluation rather than client-side training. In practice, such a validation set can be constructed from public IoT security datasets or threat-intelligence sources, making the requirement realistic for intrusion detection deployments. The current implementation of FBID uses a lightweight multilayer perceptron on tabular flow features. Since the server-side controller operates on performance and trust-related signals rather than architecture-specific gradients, the framework can in principle be extended to other model families, such as sequence-based intrusion detection architectures.

The system overhead introduced by FBID is minimal. Server-side bandit selection and validation add negligible computational time per round, while communication overhead requires transmitting only two additional scalars ($E_k$ and $\alpha_k$). Server memory requirements are, storing just a $6 \times 6$ matrix per client for LinUCB and validation is easily parallelized. Collectively, these efficiencies make the proposed control mechanism highly practical for moderate-scale federated intrusion detection.

\section{Conclusion}
In this paper, we have proposed FBID, a novel PFL framework for IoT intrusion detection. This framework employs a server-side contextual multi-armed bandit mechanism to assign client-specific local training budgets, and a trust-based blending mechanism to derive client-specific interpolation parameters between the global and local models, thereby maintaining a global knowledge floor while still allowing beneficial local specialization. Through extensive experiments on the CICIoT2023 dataset, we have shown that FBID improves the OOD DR by up to 7.66\% and the F1 by up to 5.08\% in individual client performance compared with the strongest stable baseline. For future research, the framework can be extended to asynchronous federated settings, deeper model architectures, and reduced dependence on server-side validation through more scalable supervision mechanisms.
% \section*{Acknowledgment}
% We acknowledge Ho Chi Minh City University of Technology (HCMUT), VNU-HCM for supporting this study.

\bibliographystyle{IEEEtran}
\footnotesize
% \IEEEtriggeratref{11}
\bibliography{IEEEabrv,ref}

\end{document}